\documentclass{interspeech2024}
\usepackage{booktabs}
\usepackage{multirow}
\usepackage{subcaption}
\usepackage{xcolor}
\usepackage{comment}
\usepackage{soul}

\interspeechcameraready

\title{Automatic Prediction of Amyotrophic Lateral Sclerosis Progression using Longitudinal Speech Transformer}
\name[affiliation={1}]{Liming}{Wang}
\name[affiliation={1}]{Yuan}{Gong}
\name[affiliation={1}]{Nauman}{Dawalatabad}
\name[affiliation={2}]{Marco}{Vilela}
\name[affiliation={2}]{Katerina}{Placek}
\name[affiliation={2}]{Brian}{Tracey}
\name[affiliation={2}]{Yishu}{Gong}
\name[affiliation={3}]{Alan}{Premasiri}
\name[affiliation={3}]{Fernando}{Vieira}
\name[affiliation={1}]{James}{Glass}
\address{
    $^1$MIT CSAIL, USA, $^2$Takeda Development Center Americas, Inc., USA\\
    $^3$ALS Therapy Development Institute, USA
}
\email{\liming{\href{https://github.com/cactuswiththoughts/ALST}{https://github.com/cactuswiththoughts/ALST}}{}}

\def\Figref#1{Fig.~\ref{#1}}

\def\Ls{\mathcal{L}}
\def\be{\mathbf{e}}
\def\bh{\mathbf{h}}
\def\bo{\mathbf{o}}
\def\bp{\mathbf{p}}
\def\bx{\mathbf{x}}
\def\bz{\mathbf{z}}
\def\hp{\hat{p}}
\def\hy{\hat{y}}
\def\emb{\mathrm{Embed}}
\def\lin{\mathrm{Linear}}
\def\phn{\mathrm{phn}}
\def\tra{\mathrm{Transformer}}
\def\hubert/{HuBERT}
\def\wavtovec/{wav2vec}
\def\wavtovectwo/{wav2vec 2.0}
\def\whisper/{Whisper}
\def\whispers/{whisper-small}
\def\whisperb/{whisper-base}
\def\whisperm/{whisper-medium}
\def\whisperl/{whisper-large v2}
\def\utt{\mathrm{utt}}
\newcommand{\liming}[2]{#1}{}

\begin{document}

\maketitle
 
\begin{abstract}
Automatic prediction of amyotrophic lateral sclerosis (ALS) disease progression provides a more efficient and objective alternative than manual approaches. We propose ALS longitudinal speech transformer (ALST), a neural network-based automatic predictor of ALS disease progression from longitudinal speech recordings of ALS patients. By taking advantage of high-quality pretrained speech features and longitudinal information in the recordings, our best model achieves 91.0\% AUC, improving upon the previous best model by 5.6\% relative on the ALS TDI dataset. Careful analysis reveals that ALST is capable of fine-grained and interpretable predictions of ALS progression, especially for distinguishing between rarer and more severe cases. Code is publicly available.
\end{abstract}

\section{Introduction}
Amyotrophic lateral sclerosis (ALS) is a progressive multisystem neurodegenerative disease characterized by muscular atrophy due to motor neuron degeneration and, in up to 50\% of patients, cognitive-behavioral impairment due to frontotemporal neuronal degeneration~\cite{Chio2019-als,Strong2009-als}. More than 5,000 individuals are diagnosed with ALS annually in the US, and the typical lifespan from diagnosis is 2-5 years. As yet, there is no cure for ALS and limited treatment options aiming at slowing disease progression or managing symptoms.
Precisely predicting ALS disease progression is crucial for early diagnosis and for developing new treatments. 
The standard-of-care evaluation for monitoring ALS disease progression is the ALS Functional Rating Scale - Revised~\cite{Cedarbaum1999-alsfrsr}. The ALSFRS-R uses clinician evaluation to ascertain ALS functional symptom severity across various functional domains including walking and speech. Importantly, a lower ALSFRS-R score at diagnosis as well as faster decline on the ALSFRS-R over time is linked to shorter survival from symptom onset~\cite{Kaufmann2005-als,Kollewe2008-als}.

In this paper, we examine the utility of an automated system to predict ALS disease progression from digitally-collected speech samples, inspired by previous research devoted to automatic ALS disease progression prediction~\cite{Hothorn2014-als,Gomeni2014-als,van-der-Burgh2017-als-mri,Bandini2018,Westeneng2018,Grollemund2021,Pancotti2022-als,Vieira2022-als-baseline,Jabbar2023-deeplearning-als,Gupta2023-als-data,Tavazzi2023-ml-for-als}. These systems predict patient survival rates or temporal changes in ALSFRS-R scores using machine and deep learning algorithms such as random forests~\cite{Hothorn2014-als}, dynamic Bayesian networks~\cite{Gomeni2014-als,Westeneng2018,Grollemund2021,Gupta2023-als-data} and neural networks~\cite{Pancotti2022-als,Jabbar2023-deeplearning-als}. They leverage diverse information sources from clinical documents~\cite{Hothorn2014-als,Gomeni2014-als,Westeneng2018,Grollemund2021,Pancotti2022-als,Jabbar2023-deeplearning-als} to various sensory signals such as magnetic resonance imaging scans~\cite{van-der-Burgh2017-als-mri}, motion~\cite{Bandini2018,Vieira2022-als-baseline,Gupta2023-als-data} and speech~\cite{Vieira2022-als-baseline}. The successful application of these methods could offer a more objective and consistent evaluation of ALS disease progression, potentially reducing the need for extensive human involvement in the process. 

Focusing on predicting ALS disease progression from speech has four main advantages. First, the collection of speech samples is economical and straightforward without requiring any medical expertise. This aspect makes speech-based systems not only more accessible but a cost-effective approach for tracking the ALS disease progression. Second, speech impairments are significant indicators of ALS progression. Significant speech impairment, such as lower speaking rate and intelligibility is a well-known symptom of ALS, especially the bulbar-onset cases where motor control of the tongue and lips are initially affected~\cite{Ball2002-als-speech}. Third, speech-based systems discern latent variables that can aid in the prediction of ALS disease progression.
For example, prior studies in patients with bulbar-onset ALS indicate that the physiologic biomechanics of speech production~\cite{Shellikeri2016-als-speech-movement} as well as the acoustic analysis of speech lead to increased sensitivity of motor speech impairment~\cite{Shellikeri2023-digital-markers}.
Last, recent advances in self-supervised speech representation~\cite{Baevski2020-wav2vec2,Hsu2021-hubert} and large-scale automatic speech recognition systems such as Whisper~\cite{Radford2023-whisper} have exploited the large amount of speech data online to obtain effective features that capture both linguistic and paralinguistic information in speech. Such speech features have demonstrated remarkable transferability in various related tasks such as cognitive disorder detection~\cite{Li2023-alzheimer-whisper}, pronunciation assessment~\cite{Chen2023-multipa} and audio classification~\cite{Gong2023-whisperat}. 

This paper's main contributions are threefold. First, we redefine ALS progression prediction as a longitudinal sequence prediction problem. Second, we introduce a novel model architecture that utilizes the latest pretrained speech representation models for predicting ALS disease progression on the ALSFRS-R. Lastly, we provide an extensive evaluation, examining the effects of different model features, architectural choices, and training objectives, as well as the impact of various phonemes on model predictions.

\section{Related work}
Previous studies~\cite{Vieira2022-als-baseline,Shellikeri2023-digital-markers,Gupta2023-als-data,Luz2021_alzheimer_adresso,Rohanian2021-alzheimer-speech,Balagopalan2021-alzheimer-w2v2,Li2023-alzheimer-whisper,Bowden2023-als} have shown the potential of machine learning models in predicting the progression of ALS and other neurodegenerative diseases like Alzheimer's from speech. For ALS, a multi-class classifier was trained to predict ALSFRS-R scores using speech and accelerometer data, with mel spectrograms and a convolutional neural network (CNN)~\cite{Vieira2022-als-baseline}. In addition, \cite{Shellikeri2023-digital-markers} has proposed automatic measures for ALS severity based on forced alignment and formant analysis of spontaneous speech. In Alzheimer's research, various models were developed for detection and progression prediction using different techniques such as decision trees~\cite{Luz2021_alzheimer_adresso}, recurrent neural nets~\cite{Rohanian2021-alzheimer-speech}, \wavtovectwo/~\cite{Balagopalan2021-alzheimer-w2v2} and \whisper/~\cite{Li2023-alzheimer-whisper}. However, these Alzheimer's models mainly focus on cross-sectional binary classification, unlike the longitudinal, multi-class setting in ALS research.


\section{Methods}
\subsection{Dataset}
We use the ALS Therapy Development Institute (ALS TDI) for evaluation, which is part of a ALS Research Collaborative (formerly known as the Precision Medicine Program)~\cite{Gupta2023-als-data,Vieira2022-als-baseline}. This dataset comprises speech recordings accompanied by self-assessed ALSFRS-R scores from a longitudinal study of ALS patients, primarily based in the U.S. and distributed evenly across states~\cite{Vieira2022-als-baseline}. Participants in the study were asked to repeat the sentence ``I owe you a yoyo today'' five times in each recording. Additionally, participants provided their ALSFRS-R ratings for 12 different \liming{}{cognitive and }behavioral functions at the time of recording. However, our experiments only utilize the ALSFRS-R speech sub-score. Due to missing recordings and labels and quality of text transcriptions, our experiments focus on a subset of 2,851 voice samples from 425 patients. We divide the dataset into two distinct groups: a training set with 2,124 utterances from 344 patients and a testing set with 727 utterances from 81 patients. The average years since diagnosis is 2.8 years and the average tracking period is 0.8 years with an average gap of about 50 days between successive recordings from the same patient. About 12\% and 88\% of the recordings are from bulbar-onset and limb-onset ALS patients respectively, and the average ALSFRS-R score at baseline measurement is 3.3.

\subsection{ALS transformer}
For each patient, we have a list of utterances  $[\bx^1,\cdots,\bx^n]$ labeled with self-rated ALSFRS-R scores at a 5-point scale, $[y_1,\cdots, y_n]\in \{0,\cdots,4\}^n$ and phoneme transcripts $[\mathbf{\phi}^1,\cdots,\mathbf{\phi}^n]$. Further, we also have access to the \emph{longitudinal} information of the utterances, i.e., the patient identities and recording dates $[d_1,\cdots,d_n]$ of each utterance.

\begin{figure}
    \centering
    \includegraphics[width=0.49\textwidth]{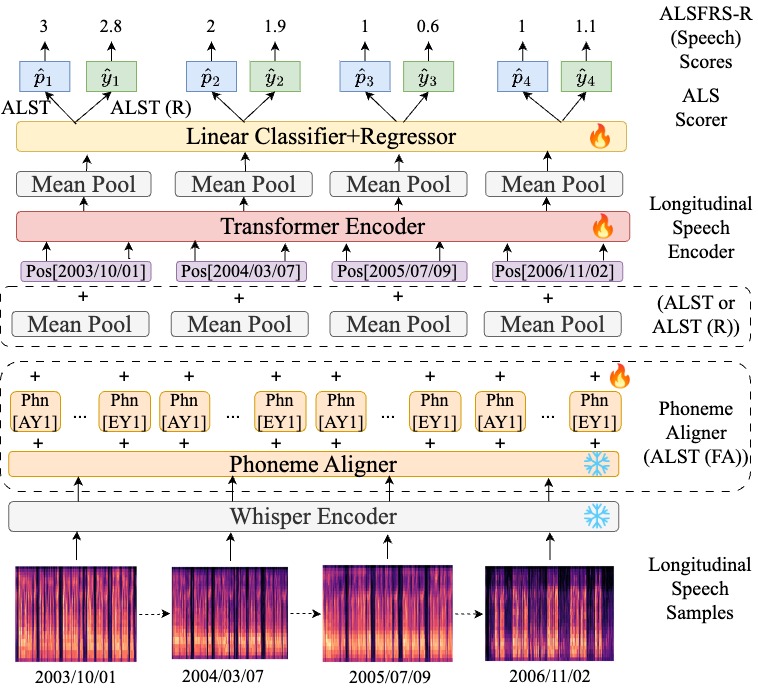}
    \caption{Architecture of the proposed ALST. The \whisper/ encoder is frozen during training.}
    \label{fig:alst}
\end{figure}

Instead of predicting each ALSFRS-R score independently, we formulate the problem as a sequence prediction problem and predict the whole self-rated ALSFRS-R scores for each patient jointly. To this end, we propose a novel architecture, ALS transformer (ALST), capable of modeling longitudinal context. As shown in \Figref{fig:alst}, ALST consists of four main components: a pretrained feature extractor, a phoneme aligner, a longitudinal speech encoder and an ALSFRS-R scorer.

The pretrained feature extractor, either a \wavtovectwo/~\cite{Baevski2020-wav2vec2} or \whisper/ encoder~\cite{Radford2023-whisper}, converts each raw speech waveform into 20ms frame-level features $[\bz^1,\cdots,\bz^n]$.  We then concatenate the speech features for all the utterances of each patient in increasing order of their recording dates and extract the phoneme-level forced alignment boundaries as \liming{$[s_1,\cdots,s_{m_1}^1,\cdots, s_1^n,\cdots,s_{m_n}^n]$}{}, where $m_i$ is the total number of phonemes in utterance $i$ and $s_j^i$ is the starting frame index of the $j$-th phoneme in utterance $i$. The phoneme aligner then converts the frame-level features to phoneme-level representation using the forced alignments:
\begin{align}
    \bz^{\phn, i}_j = \frac{1}{s_{j+1}^i-s_j^i}\sum_{s_j^i\leq t < s_{j+1}^i}\bz_t,
\end{align}
for $1\leq j\leq m_i,\,1\leq i\leq n$\liming{, with $m_i\equiv 1$ for no alignment}{}.

The longitudinal speech encoder further contextualizes the phoneme-level features by jointly encoding features across the longitudinal sequence. To this end, it first creates a sequence of trainable positional embeddings using the recording dates and a sequence of phoneme label embeddings:
\begin{align}
    \be_j^{\mathrm{pos},i} &= \emb(j)+\emb(d_{i}),\\
    \be^{\phn,i}_j &= \emb(\phi^i_j),\,1\leq j\leq m_i, 1\leq i\leq n.
\end{align}
We experiment with two main types of position embeddings for dates, namely, the \emph{order} embedding to simply encode the ranking order of the dates and the \emph{day} embedding to encode the number of days of the recording since the diagnosis date of the patient. Next, it combines the phoneme-level speech features and the embeddings using multiple standard transformer blocks:
\begin{align}
    \bar{\bh}_j^i &= \tra_j(\lin(\bz^{\phn,i}_{1:m})+\be_{1:m}^{\mathrm{pos},i})\\
    \bh_j^i &= \lin(\bar{\bh}_j^i)+\be_j^{\phn,i},\,1\leq j\leq m,\,1\leq i\leq n.
\end{align}

The ALSFRS-R scorer first converts the hidden encodings from the encoder to the utterance level via mean pooling:
\begin{align}
    \bz_i^{\utt} = \frac{1}{m_i}\sum_{j=1}^{m_i}\bh_j^i.
\end{align}
Next, it combines the scores sequence with the utterance-level features into two predictions, a continuous estimated score $\hy_i$ and an estimated probability over score classes $\hat{\bp}_i$:
\begin{align}
    [\hy_i, \bo_i] &= \lin(z_i^{\utt}),\\
    \hat{\bp}_i &= \text{Softmax}(\bo_i),\,1\leq i\leq n.
\end{align}
The training objective for ALST is then a linear combination of cross entropy and mean-squared error (MSE):
\begin{align}
    \Ls_{\text{ALST}} = \sum_{i=1}^n |\hy_i-y_i|^2 + \lambda_{\text{CE}}\log \hp_i(y_i).
\end{align}

\subsection{Preprocessing}
For the pretrained feature extractor, we use either the \wavtovectwo/ (large)~\cite{Baevski2020-wav2vec2} pretrained and finetuned on the 960-hour LibriSpeech dataset~\cite{Panayotov2015}, or the encoder output of \whisper/~\cite{Radford2023-whisper}. We have experimented with outputs from all encoder layers for both types of encoders as we will discuss in the analysis section. For the longitudinal speech encoder, we use a two-layer, single-head transformer~\cite{Vaswani2017} with 512 hidden nodes. We set $\lambda_{\text{CE}}=1$ by default and train the ALST using Adam optimizer~\cite{Kingma2015-adam} with a batch size of 32, a warmup step of 100 and an initial learning rate of $10^{-4}$ for 100 epochs.  We also use a multi-step learning rate decay starting from the 20$^{\text{th}}$ epoch with a step size of 5 and a decay rate of 0.5. On a NVIDIA RTX A5000 GPU, the training takes about 2.2 minutes. All our models has around 7 million parameters. Models from the last epoch provides all the reported test results.

\subsection{Evaluation metrics}
To evaluate the ALSFRS-R speech score classification performance of our models, we report both macro F1, i.e., one-versus-all F1 averaged over classes and accuracy. To account for inter-patient inconsistency of the self-rated ALSFRS-R speech scores and to assess how well the model captures longitudinal variation in the speech, we also report intra-patient ranking measures such as Spearman's $\rho$, Kendall's $\tau$ and the pairwise accuracy score. Using such measures, we can evaluate how well the model ranks the voice samples by the ALS severity of the patient rather than on the absolute values of the scores, which can have large variability among patients. For the pairwise score, we label $1$ a pair of scores $(y_1, y_2)$ in increasing order of their recording dates if the $y_1 > y_2$, $-1$ if $y_1 < y_2$ and $0$ otherwise. To assess how well our models provide a more fine-grained, continuous ALS progression measure. We also compute the MSE between predicted and true scores on the test set.

\section{Results}
\begin{table}[t]
    \centering
    \caption{Overall ALSFRS-R speech score prediction results.  $^*$ denotes results obtained from a larger dataset containing our dataset along with additional voice samples. \liming{ALST is used for AUC, while ALST (R) is used for other metrics}{}.}
    \label{tab:overall}
    \resizebox{0.48\textwidth}{!}{
    \begin{tabular}{lcccccc}
        \toprule
          Scorer & Macro F1$\uparrow$ & AUC$\uparrow$ & Spearman $\rho$$\uparrow$ & Kendall $\tau$$\uparrow$ & Pairwise Acc.$\uparrow$ & MSE$\downarrow$ \\
         \midrule
         \midrule
         \multicolumn{7}{c}{FBank} \\
         \midrule
         CNN~\cite{Vieira2022-als-baseline} & - & .865$^*$ & - & - & - & -\\
        \midrule
        \multicolumn{7}{c}{\wavtovectwo/}\\
        \midrule
        SVM & \begin{tabular}{@{}c@{}}\textbf{.578}\\$\pm.000$\end{tabular}& \begin{tabular}{@{}c@{}}\textbf{.906}\\$\pm.000$\end{tabular}& \begin{tabular}{@{}c@{}}.445\\$\pm.000$\end{tabular}& \begin{tabular}{@{}c@{}}.428\\$\pm.000$\end{tabular}& \begin{tabular}{@{}c@{}}.751\\$\pm.000$\end{tabular}& - \\
        \midrule
        ALST & \begin{tabular}{@{}c@{}}.544\\$\pm.075$\end{tabular}& \begin{tabular}{@{}c@{}}.900\\$\pm.011$\end{tabular} & \begin{tabular}{@{}c@{}}\textbf{.526}\\$\pm.061$\end{tabular}& \begin{tabular}{@{}c@{}}\textbf{.513}\\$\pm.060$\end{tabular}& \begin{tabular}{@{}c@{}}\textbf{.801}\\$\pm.017$\end{tabular}& \begin{tabular}{@{}c@{}}.785\\$\pm.009$\end{tabular}\\
        \midrule
         \multicolumn{7}{c}{\whisperm/}\\
         \midrule
         SVM & \begin{tabular}{@{}c@{}}.549\\$\pm.000$\end{tabular}& \begin{tabular}{@{}c@{}}\textbf{.910}\\$\pm.000$\end{tabular}& \begin{tabular}{@{}c@{}}.470\\$\pm.000$\end{tabular}& \begin{tabular}{@{}c@{}}.458\\$\pm.000$\end{tabular}& \begin{tabular}{@{}c@{}}.760\\$\pm.000$\end{tabular}& -\\
        \midrule
        ALST & \begin{tabular}{@{}c@{}}\textbf{.577}\\$\pm.054$\end{tabular}& \begin{tabular}{@{}c@{}}.906\\$\pm.007$\end{tabular}& \begin{tabular}{@{}c@{}}\textbf{.594}\\$\pm.022$\end{tabular}& \begin{tabular}{@{}c@{}}\textbf{.575}\\$\pm.024$\end{tabular}& \begin{tabular}{@{}c@{}}\textbf{.818}\\$\pm.015$\end{tabular}& \begin{tabular}{@{}c@{}}.780\\$\pm.004$\end{tabular}\\
        \midrule
        \multicolumn{7}{c}{\whisperl/}\\
        \midrule
        SVM & \begin{tabular}{@{}c@{}}.552\\$\pm.000$\end{tabular}& \begin{tabular}{@{}c@{}}\textbf{.908}\\$\pm.000$\end{tabular}& \begin{tabular}{@{}c@{}}.446\\$\pm.000$\end{tabular}& \begin{tabular}{@{}c@{}}.429\\$\pm.000$\end{tabular}& \begin{tabular}{@{}c@{}}.773\\$\pm.000$\end{tabular}& -\\
        \midrule
        ALST & \begin{tabular}{@{}c@{}}\textbf{.561}\\$\pm.032$\end{tabular}& \begin{tabular}{@{}c@{}}.907\\$\pm.013$\end{tabular}& \begin{tabular}{@{}c@{}}\textbf{.552}\\$\pm.053$\end{tabular}& \begin{tabular}{@{}c@{}}\textbf{.537}\\$\pm.053$\end{tabular}& \begin{tabular}{@{}c@{}}\textbf{.804}\\$\pm.019$\end{tabular}& \begin{tabular}{@{}c@{}}\textbf{.777}\\$\pm.011$\end{tabular}\\
        \bottomrule
    \end{tabular}
    }
\end{table}

\begin{table}[ht]
    \caption{ALSFRS-R speech score prediction results vs. different variants of ALSTs. \emph{ALST (R)} stands for the regression branch of the ALST; \emph{ALST (FA)} stands for ALST with forced aligned phoneme-level input features and outputs from the regression branch.}
    \label{tab:alst_branch}
    \resizebox{0.48\textwidth}{!}{
    \begin{tabular}{ccccccc}
        \toprule
         Variant & Macro F1$\uparrow$ & AUC$\uparrow$ & Spearman $\rho$$\uparrow$ & Kendall $\tau$$\uparrow$ & Pairwise Acc.$\uparrow$ & MSE$\downarrow$ \\
         \midrule
         \midrule
         \multicolumn{7}{c}{\wavtovectwo/}\\
         \midrule
         ALST & \begin{tabular}{@{}c@{}}.482\\$\pm.037$\end{tabular}& \begin{tabular}{@{}c@{}}.900\\$\pm.011$\end{tabular}& \begin{tabular}{@{}c@{}}.583\\$\pm.029$\end{tabular}& \begin{tabular}{@{}c@{}}.567\\$\pm.031$\end{tabular}& \begin{tabular}{@{}c@{}}.806\\$\pm.015$\end{tabular}& -\\
         \midrule
         ALST (R) & \begin{tabular}{@{}c@{}}.544\\$\pm.075$\end{tabular}& - & \begin{tabular}{@{}c@{}}.526\\$\pm.061$\end{tabular}& \begin{tabular}{@{}c@{}}.513\\$\pm.060$\end{tabular}& \begin{tabular}{@{}c@{}}.801\\$\pm.017$\end{tabular}& \begin{tabular}{@{}c@{}}.785\\$\pm.009$\end{tabular}\\
         \midrule
         ALST (FA) & \begin{tabular}{@{}c@{}}.529\\$\pm.028$\end{tabular}& \begin{tabular}{@{}c@{}}.898\\$\pm.007$\end{tabular}& \begin{tabular}{@{}c@{}}.448\\$\pm.045$\end{tabular}& \begin{tabular}{@{}c@{}}.432\\$\pm.044$\end{tabular}& \begin{tabular}{@{}c@{}}.732\\$\pm.013$\end{tabular}& \begin{tabular}{@{}c@{}}.795\\$\pm.005$\end{tabular}\\
         \midrule
         \multicolumn{7}{c}{\whisperm/}\\
         \midrule
         ALST & \begin{tabular}{@{}c@{}}.560\\$\pm.016$\end{tabular}& \begin{tabular}{@{}c@{}}.906\\$\pm.007$\end{tabular}& \begin{tabular}{@{}c@{}}.585\\$\pm.020$\end{tabular}& \begin{tabular}{@{}c@{}}.566\\$\pm.019$\end{tabular}& \begin{tabular}{@{}c@{}}\textbf{.822}\\$\pm.012$\end{tabular}& -\\
         \midrule
         ALST (R) & \begin{tabular}{@{}c@{}}.577\\$\pm.054$\end{tabular}& -& \begin{tabular}{@{}c@{}}\textbf{.594}\\$\pm.022$\end{tabular}& \begin{tabular}{@{}c@{}}\textbf{.575}\\$\pm.024$\end{tabular}& \begin{tabular}{@{}c@{}}.818\\$\pm.015$\end{tabular}& \begin{tabular}{@{}c@{}}.780\\$\pm.004$\end{tabular}\\
         \midrule
         ALST (FA) & \begin{tabular}{@{}c@{}}.571\\$\pm.069$\end{tabular}& \begin{tabular}{@{}c@{}}.907\\$\pm.009$\end{tabular}& \begin{tabular}{@{}c@{}}.469\\$\pm.070$\end{tabular}& \begin{tabular}{@{}c@{}}.451\\$\pm.067$\end{tabular}& \begin{tabular}{@{}c@{}}.731\\$\pm.012$\end{tabular}& \begin{tabular}{@{}c@{}}.783\\$\pm.010$\end{tabular}\\
         \midrule
        \multicolumn{7}{c}{\whisperl/}\\
         \midrule
         ALST & \begin{tabular}{@{}c@{}}.486\\$\pm.049$\end{tabular}& \begin{tabular}{@{}c@{}}.907\\$\pm.013$\end{tabular}& \begin{tabular}{@{}c@{}}.470\\$\pm.041$\end{tabular}& \begin{tabular}{@{}c@{}}.458\\$\pm.039$\end{tabular}& \begin{tabular}{@{}c@{}}.799\\$\pm.012$\end{tabular}& -\\
        \midrule
         ALST (R) & \begin{tabular}{@{}c@{}}.561\\$\pm.032$\end{tabular}& -& \begin{tabular}{@{}c@{}}.552\\$\pm.053$\end{tabular}& \begin{tabular}{@{}c@{}}.537\\$\pm.053$\end{tabular}& \begin{tabular}{@{}c@{}}.804\\$\pm.019$\end{tabular}& \begin{tabular}{@{}c@{}}\textbf{.777}\\$\pm.011$\end{tabular}\\
        \midrule
         ALST (FA) & \begin{tabular}{@{}c@{}}\textbf{.594}\\$\pm.035$\end{tabular}& \begin{tabular}{@{}c@{}}.909\\$\pm.005$\end{tabular}& \begin{tabular}{@{}c@{}}.520\\$\pm.049$\end{tabular}& \begin{tabular}{@{}c@{}}.502\\$\pm.048$\end{tabular}& \begin{tabular}{@{}c@{}}.753\\$\pm.027$\end{tabular}& \begin{tabular}{@{}c@{}}.780\\$\pm.009$\end{tabular}\\
        \bottomrule
    \end{tabular}
    }
\end{table}

\begin{table}[ht]
    \caption{ALSFRS-R speech score prediction results vs. different setups for ALST (R)s with \whisperm/ encoders. The setup ``longitudinal'' uses longitudinal sequence without positional embeddings.}
    \resizebox{0.48\textwidth}{!}{
    \begin{tabular}{ccccccc}
        \toprule
         Setup & Macro F1$\uparrow$ & AUC$\uparrow$ & Spearman $\rho$$\uparrow$ & Kendall $\tau$$\uparrow$ & Pairwise Acc.$\uparrow$ & MSE$\downarrow$ \\
         \midrule
         \midrule
         No longitudinal & \begin{tabular}{@{}c@{}}\textbf{.581}\\$\pm.017$\end{tabular}& \begin{tabular}{@{}c@{}}\textbf{.916}\\$\pm.003$\end{tabular}& \begin{tabular}{@{}c@{}}.546\\$\pm.081$\end{tabular} & \begin{tabular}{@{}c@{}}.529\\$\pm.080$\end{tabular} & \begin{tabular}{@{}c@{}}.796\\$\pm.022$\end{tabular} & \begin{tabular}{@{}c@{}}\textbf{.768}\\$\pm.005$\end{tabular}\\
         \midrule
         Longitudinal & \begin{tabular}{@{}c@{}}.552\\$\pm.008$\end{tabular}& \begin{tabular}{@{}c@{}}.912\\$\pm.010$\end{tabular}& \begin{tabular}{@{}c@{}}\textbf{.611}\\$\pm.046$\end{tabular}& \begin{tabular}{@{}c@{}}\textbf{.592}\\$\pm.045$\end{tabular}& \begin{tabular}{@{}c@{}}\textbf{.826}\\$\pm.015$\end{tabular}& \begin{tabular}{@{}c@{}}.772\\$\pm.009$\end{tabular}\\
         \midrule
         Order, trainable & \begin{tabular}{@{}c@{}}.558\\$\pm.017$\end{tabular}& \begin{tabular}{@{}c@{}}.914\\$\pm.011$\end{tabular}& \begin{tabular}{@{}c@{}}.564\\$\pm.083$\end{tabular}& \begin{tabular}{@{}c@{}}.543\\$\pm.081$\end{tabular}& \begin{tabular}{@{}c@{}}.813\\$\pm.014$\end{tabular}& \begin{tabular}{@{}c@{}}.774\\$\pm.009$\end{tabular}\\
         \midrule
         Order, sinusoid & \begin{tabular}{@{}c@{}}.520\\$\pm.028$\end{tabular}& \begin{tabular}{@{}c@{}}.912\\$\pm.009$\end{tabular}& \begin{tabular}{@{}c@{}}.561\\$\pm.107$\end{tabular}& \begin{tabular}{@{}c@{}}.543\\$\pm.106$\end{tabular}& \begin{tabular}{@{}c@{}}.805\\$\pm.022$\end{tabular}& \begin{tabular}{@{}c@{}}.771\\$\pm.006$\end{tabular}\\
        \midrule
        Day, sinusoid & \begin{tabular}{@{}c@{}}.536\\$\pm.023$\end{tabular}& \begin{tabular}{@{}c@{}}.903\\$\pm.005$\end{tabular}& \begin{tabular}{@{}c@{}}.522\\$\pm.019$\end{tabular}& \begin{tabular}{@{}c@{}}.506\\$\pm.018$\end{tabular}& \begin{tabular}{@{}c@{}}.803\\$\pm.017$\end{tabular}& \begin{tabular}{@{}c@{}}.782\\$\pm.006$\end{tabular}\\
        \bottomrule
    \end{tabular}
    }
\end{table}
The overall results of our models are shown in Table~\ref{tab:overall}. We compare ALST with two comparator baseline models: a CNN-based model with mel filterbank (FBank) features as inputs~\cite{Vieira2022-als-baseline} and a linear SVM with the same speech features as the ALST. Models using pretrained speech representations consistently outperform the FBank-based baseline by around 5.6\% relative AUC. 
Among the pretrained speech representations, \whisperm/ encoder performs the best in all the ranking-based metrics, even better than the larger \whisperl/ encoder in all except the macro F1 score. While this might be due to training instability or scaling issues for \whisperl/, another possible explanation is that \whisperl/ has filtered out some of the paralinguistic information due to their irrelevance to its original task of ASR.

Our experiments also demonstrate the advantage of nonlinear models such as transformers~\cite{Vaswani2017} over linear models for ALS prediction. With the same speech representation, the best ALSTs generally outperforms the linear SVMs in all comparable metrics, especially ranking-based metrics, with around 5\% relative improvements in pairwise accuracy and around 25\% relative improvements in both Spearman $\rho$ and Kendall $\tau$. For the Macro F1, the \whisper/-based ALSTs outperform the SVM by an average of 6.5\% relative, while the \wavtovectwo/-based ALSTs performs worse than the SVM by an 5.9\% relative.
Moreover, as shown in Table~\ref{tab:alst_branch} the regression branch of ALST consistently outperforms the classification branch in macro F1 for all pretrained speech representations, as well as all the ranking-based metrics for whisper representations. The use of phoneme segmentation helps \whisperl/ in Macro F1 and AUC, but not ranking-based metrics or with other pretrained speech representations.
\begin{figure}[ht]
    \centering
    \begin{subfigure}{0.23\textwidth}
    \includegraphics[width=0.99\textwidth]{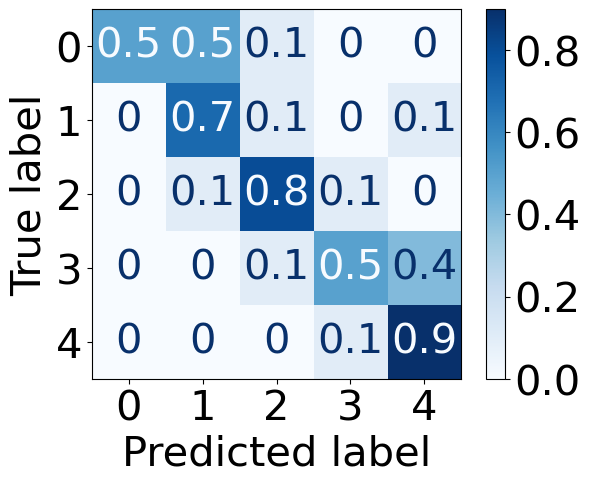}
    \caption{}
    \label{fig:confusion_alst}
    \end{subfigure}
    \begin{subfigure}{0.23\textwidth}
    \centering
    \includegraphics[width=0.99\textwidth]{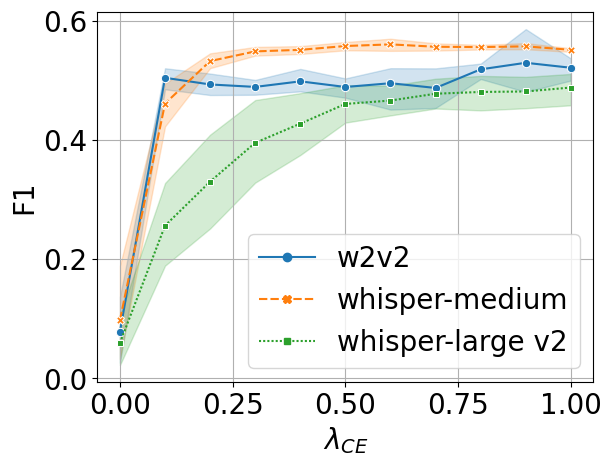}
    \caption{}
    \label{fig:eff_ce_weight}
    \end{subfigure}
    \begin{subfigure}{0.3\textwidth}
    \centering
    \includegraphics[width=0.99\textwidth]{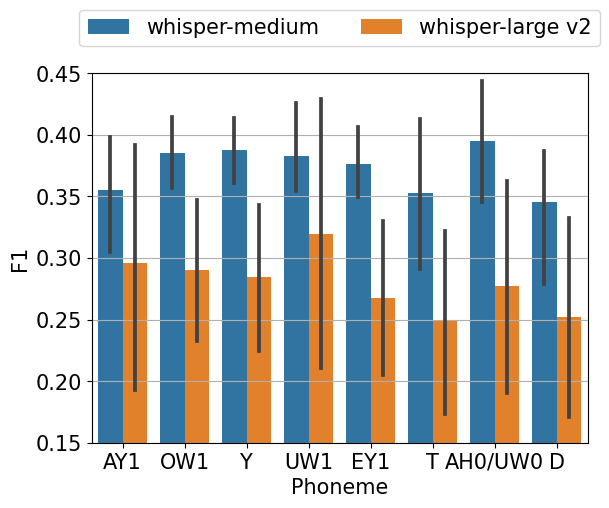}
    \caption{}
    \label{fig:eff_of_phonemes}
    \end{subfigure}
    \caption{(a) Confusion matrix of ALST (R) trained using \whisperm/ speech encoders (layer 22); (b) Effect of $\lambda_{\text{CE}}$ on ALSFRS-R speech score prediction performance; (c) Single-phoneme ALS prediction F1 vs phoneme used for the ALST models with different pretrained speech representations. We merge results for the phonemes `AH0' and `UW0' due to the multiple possible pronunciations in the second phoneme of the word `t\textbf{o}day'.}
\end{figure}

\begin{figure}[ht]
    \centering
    \begin{subfigure}{0.49\textwidth}
    \includegraphics[width=0.99\textwidth]{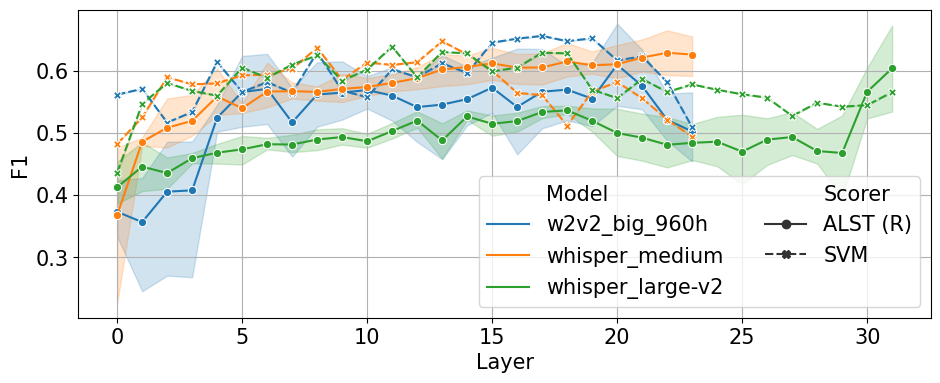}
    \caption{Macro F1 vs. layer of the pretrained model}
    \label{fig:f1_vs_layer}
    \end{subfigure}
    \begin{subfigure}{0.49\textwidth}
    \includegraphics[width=0.99\textwidth]{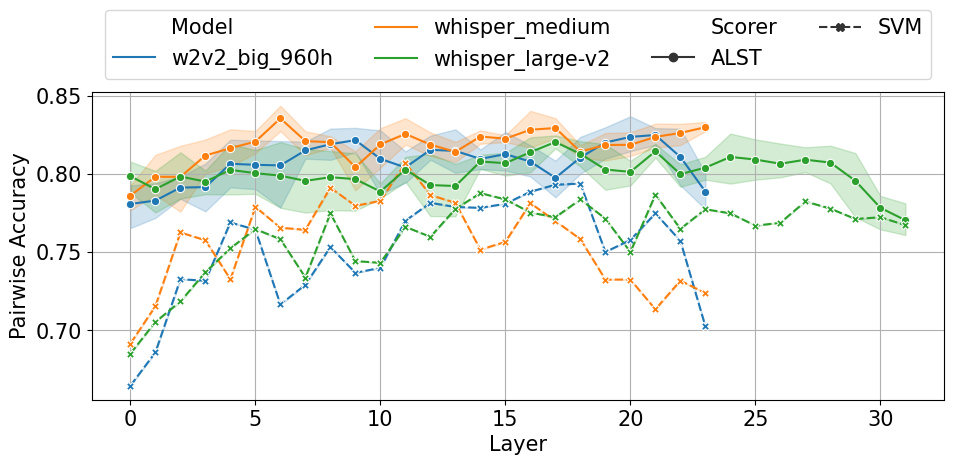}
    \caption{Pairwise accuracy vs. layer of the pretrained model}
    \label{fig:pw_acc_vs_layer}
    \end{subfigure}
    \caption{Score prediction performance vs. layer used for feature extraction of the pretrained speech representation models}
    \label{fig:result_vs_layer}
\end{figure}

To better understand the role of longitudinal information for ALS prediction, we experiment with different setups including using no longitudinal information, using longitudinal information without position embeddings and with different types of position embeddings. Our experiments indicate that longitudinal information significantly improves ranking-based metrics, but not classification metrics. The results demonstrate that the two types of metrics do measure different aspects of ALS prediction performance, and that longitudinal information helps ALST to better estimate the disease progression for a single speaker, which is closer to our goal than to predict the absolute values of the ALS speech scores due to inter-speaker variability in rating standards. However, we observe both the order-based position embeddings and the day-based position embeddings degrades the performance obtained using longitudinal information without position embeddings. Designing a better position embedding for longitudinal ALS prediction will be a future research direction. 

We also visualize the confusion matrix generated by our best model in terms of macro F1 in \Figref{fig:confusion_alst}. ALST performs the best for recordings with a ALSFRS-R speech score of 4, the most frequent score class while the worst for score of 0, the rarest score class. Interestingly, most mistakes made by ALST are between similar score classes such as 0 vs. 1 and 3 vs. 4, many of which may be due to individual differences in rating standards among patients. Therefore, macro F1 potentially \emph{underestimates} the ability of ALST for ALS prediction. Another ablation study is on the effect of relative weights between the classification (CE) and the regression (MSE) losses, shown in \Figref{fig:eff_ce_weight}. Our results suggest that training the model with MSE alone leads to poor performance, and combining CE and MSE losses with $\lambda_{CE} \approx 0.9$ leads the best result. The benefit of using the CE loss saturates after $\lambda_{CE}$ exceeds around 0.6. In \Figref{fig:eff_of_phonemes}, we estimate the importance of a given phoneme in ALS prediction. we leave one segment of that phoneme intact and mask out the rest of the recording as input to a trained ALST (FA) system, and measure the macro F1 of its prediction.  While we observe high variance across the phonemes, we can see that vowels such as ``UW1'' and ``OW1'' lead to higher macro F1s than consonants such as ``T'' and ``D''.

The final ablation concerns the importance of pretrained speech representations from different encoder layers, we train an ALST and a linear SVM for each layer of the pretrained speech encoders, as shown in \Figref{fig:result_vs_layer}. For F1 score (\Figref{fig:f1_vs_layer}), later layers generally play a bigger role in all three pretrained encoders, whereas for pairwise accuracy (\Figref{fig:pw_acc_vs_layer}), earlier layers play a bigger role.

\section{Conclusions}
This paper introduces ALST, a speech-based automatic system for predicting ALS disease progression. ALST stands out for its ability to learn  from longitudinal speech data, enabling hassle-free, long-term monitoring tailored to individual ALS patients. Notably, ALST surpasses existing speech-based system by a 5.6\% increase in relative AUC score. Future works include a more in-depth comparison with other ALS disease progression models and enhanced ways to process and interpret longitudinal speech data.


\section{Acknowledgements}
This research was supported by Takeda Development Center Americas, INC. (successor in interest to Millennium Pharmaceuticals, INC.)  This research could not be possible without the immense contributions from people living with ALS who have participated for months and years in the ALS Research Collaborative Study. We thank the ALS patient and caregiver community, in particular Augie’s Quest for a Cure, for financially supporting ALS TDI's data collection work. 


\bibliographystyle{IEEEtran}
\bibliography{reference}

\begin{thebibliography}{10}
\providecommand{\url}[1]{#1}
\csname url@samestyle\endcsname
\providecommand{\newblock}{\relax}
\providecommand{\bibinfo}[2]{#2}
\providecommand{\BIBentrySTDinterwordspacing}{\spaceskip=0pt\relax}
\providecommand{\BIBentryALTinterwordstretchfactor}{4}
\providecommand{\BIBentryALTinterwordspacing}{\spaceskip=\fontdimen2\font plus
\BIBentryALTinterwordstretchfactor\fontdimen3\font minus
  \fontdimen4\font\relax}
\providecommand{\BIBforeignlanguage}[2]{{%
\expandafter\ifx\csname l@#1\endcsname\relax
\typeout{** WARNING: IEEEtran.bst: No hyphenation pattern has been}%
\typeout{** loaded for the language `#1'. Using the pattern for}%
\typeout{** the default language instead.}%
\else
\language=\csname l@#1\endcsname
\fi
#2}}
\providecommand{\BIBdecl}{\relax}
\BIBdecl

\bibitem{Chio2019-als}
A.~Chi{\`o} \emph{et~al.}, ``{Cognitive impairment across ALS clinical stages
  in a population-based cohort},'' \emph{Neurology}, vol.~93, no.~10, pp.
  984--994, 2019.

\bibitem{Strong2009-als}
M.~J. Strong \emph{et~al.}, ``Consensus criteria for the diagnosis of
  frontotemporal cognitive and behavioural syndromes in amyotrophic lateral
  sclerosis,'' \emph{Amyotroph Lateral Scler}, vol.~10, no.~3, pp. 131--146,
  2009.

\bibitem{Cedarbaum1999-alsfrsr}
J.~M. Cedarbaum \emph{et~al.}, ``{The ALSFRS-R: a revised {ALS} functional
  rating scale that incorporates assessments of respiratory function. BDNF ALS
  Study Group (Phase III)},'' \emph{J. Neurol. Sci.}, vol. 169, pp. 13--21,
  1999.

\bibitem{Kaufmann2005-als}
\BIBentryALTinterwordspacing
P.~Kaufmann, G.~Levy, J.~L. Thompson, M.~L. DelBene, V.~Battista, P.~H. Gordon,
  L.~P. Rowland, B.~Levin, and H.~Mitsumoto, ``{The ALSFRS-R predicts survival
  time in an ALS clinic population},'' \emph{Neurology}, vol.~64, no.~1, pp.
  38--43, 2005. [Online]. Available:
  \url{https://www.neurology.org/doi/abs/10.1212/01.WNL.0000148648.38313.64}
\BIBentrySTDinterwordspacing

\bibitem{Kollewe2008-als}
\BIBentryALTinterwordspacing
K.~Kollewe, U.~Mauss, K.~Krampfl, S.~Petri, R.~Dengler, and B.~Mohammadi,
  ``{ALSFRS-R score and its ratio: A useful predictor for ALS-progression},''
  \emph{Journal of the Neurological Sciences}, vol. 275, no. 1-2, pp. 69--73,
  2008. [Online]. Available: \url{https://doi.org/10.1016/j.jns.2008.07.016}
\BIBentrySTDinterwordspacing

\bibitem{Hothorn2014-als}
T.~Hothorn and H.~H. Jung, ``{RandomForest4Life: A Random Forest for predicting
  ALS disease progression},'' \emph{Amyotrophic Lateral Sclerosis and
  Frontotemporal Degeneration}, vol.~15, no. 5-6, pp. 444--452, 2014.

\bibitem{Gomeni2014-als}
R.~Gomeni and M.~Fava, ``Amyotrophic lateral sclerosis disease progression
  model,'' \emph{Amyotrophic Lateral Sclerosis and Frontotemporal
  Degeneration}, vol.~15, no. 1-2, pp. 119--129, 2014.

\bibitem{van-der-Burgh2017-als-mri}
{Hannelore K. van der Burgh and Ruben Schmidt and Henk-Jan Westeneng and Marcel
  A. de Reus and Leonard H. van den Berg and Martijn P. van den Heuvel},
  ``{Deep learning predictions of survival based on MRI in amyotrophic lateral
  sclerosis},'' \emph{NeuroImage: Clinical}, vol.~13, pp. 361--369, 2017.

\bibitem{Bandini2018}
\BIBentryALTinterwordspacing
A.~Bandini \emph{et~al.}, ``Kinematic features of jaw and lips distinguish
  symptomatic from presymptomatic stages of bulbar decline in amyotrophic
  lateral sclerosis,'' \emph{J Speech Lang Hear Res}, vol.~61, no.~5, pp.
  1118--1129, 2018. [Online]. Available:
  \url{https://pubs.asha.org/doi/10.1044/2018_JSLHR-S-17-0262}
\BIBentrySTDinterwordspacing

\bibitem{Westeneng2018}
H.-J. Westeneng \emph{et~al.}, ``Prognosis for patients with amyotrophic
  lateral sclerosis: Development and validation of a personalised prediction
  model,'' \emph{Lancet Neurol}, vol.~17, pp. 423--433, 2018.

\bibitem{Grollemund2021}
V.~Grollemund \emph{et~al.}, ``Manifold learning for amyotrophic lateral
  sclerosis functional loss assessment: Development and validation of a
  prognosis model,'' \emph{J Neurol}, vol. 268, no.~3, pp. 825--850, 2021.

\bibitem{Pancotti2022-als}
\BIBentryALTinterwordspacing
C.~Pancotti \emph{et~al.}, ``Deep learning methods to predict amyotrophic
  lateral sclerosis disease progression,'' \emph{Sci Rep}, vol.~12, no.~1, p.
  13738, 2022. [Online]. Available:
  \url{https://www.nature.com/articles/s41598-022-17805-9}
\BIBentrySTDinterwordspacing

\bibitem{Vieira2022-als-baseline}
\BIBentryALTinterwordspacing
F.~Vieira, S.~Venugopalan, A.~Premasiri \emph{et~al.}, ``A machine-learning
  based objective measure for {ALS} disease severity,'' \emph{{NPJ Digital
  Medicine}}, vol.~5, p.~45, 2022. [Online]. Available:
  \url{https://doi.org/10.1038/s41746-022-00588-8}
\BIBentrySTDinterwordspacing

\bibitem{Jabbar2023-deeplearning-als}
M.~A. D.~A. Jabbar \emph{et~al.}, ``{Predicting amyotrophic lateral sclerosis
  (ALS) progression with machine learning},'' \emph{Amyotrophic Lateral
  Sclerosis and Frontotemporal Degeneration}, vol.~0, no.~0, pp. 1--14, 2023.

\bibitem{Gupta2023-als-data}
\BIBentryALTinterwordspacing
A.~S. Gupta, S.~Patel, A.~Premasiri \emph{et~al.}, ``At-home wearables and
  machine learning sensitively capture disease progression in amyotrophic
  lateral sclerosis,'' \emph{Nature Communications}, vol.~14, p. 5080, 2023.
  [Online]. Available: \url{https://doi.org/10.1038/s41467-023-40917-3}
\BIBentrySTDinterwordspacing

\bibitem{Tavazzi2023-ml-for-als}
\BIBentryALTinterwordspacing
E.~Tavazzi \emph{et~al.}, ``Artificial intelligence and statistical methods for
  stratification and prediction of progression in amyotrophic lateral
  sclerosis: A systematic review,'' \emph{Artificial Intelligence in Medicine},
  vol. 142, p. 102588, 2023. [Online]. Available:
  \url{https://www.sciencedirect.com/science/article/pii/S0933365723001021}
\BIBentrySTDinterwordspacing

\bibitem{Ball2002-als-speech}
L.~J. Ball, ``Timing of speech deterioration in people with amyotrophic lateral
  sclerosis,'' \emph{Journal of Medical Speech-Language Pathology}, vol.~10,
  pp. 231--235, 2002.

\bibitem{Shellikeri2016-als-speech-movement}
S.~Shellikeri, J.~R. Green, M.~Kulkarni, P.~Rong, R.~Martino, L.~Zinman, and
  Y.~Yunusova, ``Speech movement measures as markers of bulbar disease in
  amyotrophic lateral sclerosis,'' \emph{Journal of Speech, Language, and
  Hearing Research}, vol.~59, no.~5, pp. 887--899, 2016.

\bibitem{Shellikeri2023-digital-markers}
S.~Shellikeri \emph{et~al.}, ``Digital markers of motor speech impairments in
  spontaneous speech of patients with als-ftd spectrum disorders,''
  \emph{Amyotroph Lateral Scler Frontotemporal Degener}, vol. 2023, no.
  December 5, pp. 1--9, 2023.

\bibitem{Baevski2020-wav2vec2}
A.~Baevski, H.~Zhou, A.~Mohamed, and M.~Auli, ``wav2vec 2.0: A framework for
  self-supervised learning of speech representations,'' in \emph{NeurIPS},
  2020.

\bibitem{Hsu2021-hubert}
W.-N. Hsu, B.~Bolte, Y.-H.~H. Tsai, K.~Lakhotia, R.~Salakhutdinov, and
  A.~Mohamed, ``{HuBERT}: Self-supervised speech representation learning by
  masked prediction of hidden units,'' \emph{IEEE/ACM Trans. Audio, Speech and
  Lang. Proc.}, vol.~29, pp. 3451--3460, 2021.

\bibitem{Radford2023-whisper}
A.~Radford, J.~W. Kim, T.~Xu, G.~Brockman, C.~McLeavey, and I.~Sutskever,
  ``Robust speech recognition via large-scale weak supervision,'' in
  \emph{ICML}, 2023.

\bibitem{Li2023-alzheimer-whisper}
\BIBentryALTinterwordspacing
J.~Li, K.~Song, J.~Li, B.~Zheng, D.~Li, X.~Wu, X.~Liu, and H.~Meng,
  ``Leveraging pretrained representations with task-related keywords for
  alzheimer's disease detection,'' in \emph{ArXiv}, 2023. [Online]. Available:
  \url{https://arxiv.org/pdf/2303.08019.pdf}
\BIBentrySTDinterwordspacing

\bibitem{Chen2023-multipa}
Y.-W. Chen, Z.~Yu, and J.~Hirschberg, ``{multiPA}: A multi-task speech
  pronunciation assessment system for a closed and open response scenario,'' in
  \emph{Interspeech}, 2023.

\bibitem{Gong2023-whisperat}
Y.~Gong, S.~Khurana, L.~Karlinsky, and J.~Glass, ``{Whisper-AT}: Noise-robust
  automatic speech recognizers are also strong audio event taggers,'' in
  \emph{Proc. Interspeech 2023}, 2023.

\bibitem{Luz2021_alzheimer_adresso}
S.~Luz, F.~Haider, S.~de~la Fuente, D.~Fromm, and B.~MacWhinney, ``Detecting
  cognitive decline using speech only: The {ADReSSo} challenge,'' in
  \emph{Interspeech}, 2021, pp. 3780--3784.

\bibitem{Rohanian2021-alzheimer-speech}
M.~Rohanian, J.~Hough, and M.~Purver, ``Alzheimer’s dementia recognition
  using acoustic, lexical, disfluency and speech pause features robust to noisy
  inputs,'' in \emph{Interspeech}, 2021, pp. 3820--3824.

\bibitem{Balagopalan2021-alzheimer-w2v2}
A.~Balagopalan and J.~Novikova, ``Comparing acoustic-based approaches for
  alzheimer’s disease detection,'' in \emph{Interspeech}, 2021, pp.
  3800--3804.

\bibitem{Bowden2023-als}
\BIBentryALTinterwordspacing
M.~Bowden \emph{et~al.}, ``A systematic review and narrative analysis of
  digital speech biomarkers in motor neuron disease,'' \emph{NPJ Digit Med},
  vol.~6, no.~1, p. 228, 2023. [Online]. Available:
  \url{https://www.nature.com/articles/s41746-023-00959-9}
\BIBentrySTDinterwordspacing

\bibitem{Panayotov2015}
P.~Vassil, G.~Chen, D.~Povey, and S.~Khudanpur, ``Librispeech: an {ASR} corpus
  based on public domain audio books,'' in \emph{ICASSP}, 2015, p. pp.
  5206–5210.

\bibitem{Vaswani2017}
Vaswani \emph{et~al.}, ``Attention is all you need,'' in \emph{NeurIPS}, 2017,
  p. 6000–6010.

\bibitem{Kingma2015-adam}
\BIBentryALTinterwordspacing
D.~P. Kingma and J.~L. Ba, ``Adam: A method for stochastic optimization,'' in
  \emph{ICLR}, 2015. [Online]. Available:
  \url{https://arxiv.org/pdf/1412.6980.pdf}
\BIBentrySTDinterwordspacing

\end{thebibliography}

\end{document}